\documentclass[11pt]{article}

\usepackage{natbib}
\include{graphics}
\usepackage{amsmath,epsfig,epstopdf}

\newtheorem{theorem}{Theorem}[section]

\newtheorem{corollary}{Corollary}[section]
\newenvironment{proof}{{\bf Proof:}}

\title{On a stochastic model of epidemic spread with an application to competing infections}

\author{\AA ke Svensson}
\date{May 2018}

\begin{document}

\maketitle

\begin{abstract}
A simple, but ``classical``, stochastic model for epidemic spread in a finite, but large, population is studied. The progress of the epidemic can be divided into three different phases that  requires different tools to analyse. Initially the process is approximated by a branching process. It is discussed for how long time this approximation is valid. When a non-negligible proportion of the population is already infected the process can be studied using differential equations. In a final phase the spread will fade out. The results are used to investigate what happens if two strains of  infectious agents, with different potential for spread, are simultaneously  introduced in a totally susceptible population. It is assumed that an infection causes immunity, and that a person can only be infected by one strain. The two epidemics will initially develop approximately as independent branching processes. However, if both strains causes large epidemics they will, due to  immunity, eventually interact.  We will mainly be interested in the final outcome of the spread, i.e., how large proportion of the population is infected by the different strains.
\end{abstract}


\vfill
\pagebreak

\section{Introduction}

We will consider a  SEIR-model for epidemic spread of infections in a large closed population.  The model is a stochastic version of a deterministic model studied by \cite{KMK27}.

 First we investigate what may happen when a single strain of an infectious agent is introduced in a totally susceptible population. The model is described in section \ref{model} where basic notation and important, and well-known properties, of the model are summarized.

The progress of the epidemic can be divided into three phases. As long as there is a small proportion of  immune persons in the population the spread can be approximated by a branching process. This approximation is studied in section \ref{BP}. When a non-negligible proportion of the population has been infected, if this ever happens, the branching process approximation is no longer applicable. In this phase  there are many active spreaders and the process can be analysed by mass-action tools, i.e., differential equations. This is discussed in section \ref{EP}. Finally the epidemic enters a fading off phase and will slowly die out, see section \ref{FP}.

The results obtained for the spread of a single infectious agent are used in section \ref{CE} to give  a heuristic argument of what happens if two different agents, or strains of an infectious agent, are introduced simultaneously. It is assumed that infection by one strain causes immunity to infection of the other. \cite{svetom} give a rigorous treatment valid when there are no latent periods  and the durations of the infectious periods are exponentially distributed. In this paper more general models are considered.

\section{Model for a single infection}\label{model}

At time $t=0$, one recently infected person enters into a finite totally susceptible population.  This may start a chain of infections that eventually causes an epidemic. The  spread, both as regards how large proportion of the population that are infected and the speed at which the infection grows, will depend on the infectiousness of the infected individuals. 

We will use a simplistic model for the infectiousness. After a, possibly random latent time an infected individual is infectious during a, possible random infectious time. During this time the infected has infectious contacts with randomly chosen members of the population according to a homogeneous Poisson process, with intensity $\lambda$. A contact between an infected individual and a susceptible individual will cause a secondary infection. An individual may only be infected once, i.e. is immune after infection.  The progress of the infectiousness of different infected individuals, i.e. their latent and infectious times, are assumed to be random and independent.

We will assume that the population has $n$ members, who initially are all susceptible to infection. The progress of the epidemic is followed by counting the number of infected individuals. The counting process, $N(t)$,  tells how many individuals have been infected up till time $t$. We will focus on results that are asymptotically valid when $n$ is large. 

A formal description of the epidemic is as follows:
Let $Y_i$ be the latent time and $X_i$ the infectious time of the $i$'th infected. The corresponding times of the individual introducing the infection into the population are denoted by subindex $0$. The pairs $(Y_i,X_i)$ $i=0,\ldots n$, are assumed to be independent.

Let
\begin{equation}\label{infect} 
r_i(t) = \left\{ \begin{array}{ll}
0 & \mbox{if $t<Y_i$}\\
1 & \mbox{if $Y_i\leq t < Y_i+X_i$}\\
0 & \mbox{if $t\geq Y_i+X_i$}
\end{array}
\right.
\end{equation}
$r_i(t)=1$ if the $i$'th infected is infectious at time $t$ after being infected and $0$ otherwise.

The intensity of the counting process $N$ is

\begin{equation}\label{intensity}
\lambda\left(1-\frac{N(t-)}{n}\right) \left[ \int\limits_0^t r_{N(s)}(t-s)dN(s) + r_0(t)\right].
\end{equation}

\subsection{Notations and relations}

We will use the following notations (some already introduced):

\begin{itemize}
\item $n$: The size of the population.

\item $N(t)$: The number of infected up till time $t$.

\item $Y$: A random variable describing the (individual) latent time. The density is denoted by $f_Y$ and the distribution function by $F_Y$. The Laplace transform of $Y$ is $L_Y$. For simplicity we assume that the distribution is continous and has moments of sufficiently high order,

\item $X$: A random variable describing the (individual) infectious time.  The density is denoted by $f_X$ and the distribution function by $F_X$. The Laplace transform of $X$ is $L_X$.  For simplicity we assume that the distribution is continous and has moments of sufficiently high order,

\item $F_{XY}$: The simultaneous distribution for $X$ and $Y$. Often it is assumed  that $X$ and $Y$ are  independent. This makes some of the derivations simpler but here we will not generally use this assumption.
\end{itemize}
\begin{itemize}

\item ${\lambda}$: A constant describing the intensity at which an individual takes contacts in the population.

\item $R_0$: The basic reproduction number. 
\begin{equation}
R_0 = \lambda{\rm E}(X).
\end{equation}

\item $p$: The probability that the epidemic grows large asymptotically. This probability is positive if and only if $R_0 > 1$. It is then the positive solution of the equation:
\begin{equation}\label{small}
1-p = L_X(\lambda p).
\end{equation}
With probability $q=1-p$, the introduction of the infection in the population will only cause a few infections. With probability $p$ a positive proportion of the population will finally be infected. If we later event happens we will say that the epidemic grows large, or that the epidemic has a mayor outbreak.

\item $\pi$: The final size of the epidemic, i.e. the proportion of the population that gets infected if the epidemic grows large. If the population is large and $R_0 > 1$, $\pi$ is the positive solution of the equation
\begin{equation}\label{final}
-\ln(1-\pi) = R_0\pi.
\end{equation}

\item $g$: The generation time density equals (see \cite{Svens1})
\begin{equation}\label{gdef}
g(t)=\frac{{\rm P}(Y<t<X+Y)}{{\rm E}(X)}.
\end{equation}
The corresponding distribution function is 
\begin{equation}
G(t)=\int\limits_0^tg(t)dt.
\end{equation}
The Laplace transform of $g$ equals
\begin{equation}
L_g(s)=\frac{L_Y(s)-L_{X+Y}(s)}{s{\rm E}(X)}.
\end{equation}
In case $Y$ and $X$ are independent:
\begin{equation}
L_g(s) =\frac{L_Y(s)(1-L_X(s))}{s{\rm E}(X)}.
\end{equation}

\item $T$: The mean generation time
\begin{equation}
T=\int\limits_0^{\infty} tg(t)dt = -L'_g(0).
\end {equation}

\item $\alpha$: The Malthusian parameter  $\alpha$ is the positive solution of the equation
\begin{equation}\label{me}
1=R_0\int\limits_0^{\infty}\exp(-\alpha t)g(t)dt=R_0 L_g(\alpha).
\end{equation}
or equivalently
\begin{equation}\label{al}
\alpha=\lambda \left(L_Y(\alpha)-L_{X+Y}(\alpha)\right).
\end{equation}
There will exist a (unique) positive solution to this equation if and only if $R_0>1$.

\end{itemize}

It is worth noting that different models can have the same generation time density (see \cite{svens2}). In fact, for any model, there  exists a model with independent latent and infectious times with the same generation time density. To see this, observe that

\begin{equation}
\tilde L(s)=\frac{L_Y(s)-L_{X+Y}(s)}{1-L_X(s)}
\end{equation}
is a positive completely monotone function with $
\tilde L(0)=1$. Thus it is the Laplace transform of a random variable $\tilde Y$ and

\begin{equation}
L_g(s) = \frac{\tilde L(s)(1-L_X(s))}{s{\rm E(X)}}.
\end{equation}
This implies that $g$ is the generation time density for a model with latent time distributed as $\tilde Y$ and infectious time distribution $X$ where $\tilde Y$ and $X$ are independent.
\vskip 1cm

Of course, there are intrinsic relations between the parameters. E.g.

\begin{theorem}\label{alfa}
\begin{equation}\label{kvot}
\frac{p\lambda}{\alpha}\geq 1
\end{equation}
with equality if and only if $Y\equiv 0$, i.e. when there is no latent time. 
\end{theorem}

{\bf Proof:}  By combining (\ref{small}) and (\ref{al}) when $Y\equiv 0$, and thus $L_Y(s)\equiv 1$, we find that in this case
$p=\alpha/\lambda$.

 Let
\begin{equation}
V(s) = 1-\lambda\frac{L_Y(s)-L_{X+Y}(s)}{s}
\end{equation}
$V$ is a non-decreasing function with $V(\infty)=1$. $V(0)<0$ if $R_0>1$. Thus there exist a unique positive solution, $\alpha$, which is the Malthusian parameter, such that $V(\alpha)=0$. Now let
\begin{equation}
\tilde V(s)= 1-\lambda\frac{1-L_{X}(s)}{s}
\end{equation}
The equation $\tilde V(s) =0$ has a unique positive solution, $\tilde \alpha$. Observe that $p$ depends only on $\lambda$ and $L_X$. Thus
\begin{equation}
\frac{p\lambda}{\tilde \alpha} =1.
\end {equation}
Now
\begin{equation}
V(s)-\tilde V(s) = \lambda\frac{\left[{\rm E}((1-e^{-sX})(1-e^{-sY}))\right]}{s} \geq 0,
\end{equation}
which implies
\begin{equation}
\alpha \leq \tilde \alpha.
\end{equation}
The theorem follows from this inequality.

\subsection{The Phases of the epidemic}

If the population in which the epidemic takes place is large we can divide the progress of the epidemic  into three phases. Each phase has to be analyzed with different methods.

 In the first phase it is a small probability that infected individuals will have contact with an already immune person. The progress is then not slowed down by the possibility of immunity and the epidemic can  be approximated by a branching process. Such processes have been studied in great detail. In section \ref{BP} we give a short account of  results that are relevant for the present study.
 
 If the epidemic reach a level were a non-negligible proportion of the population is infected it will enter into a mass-action phase where the progress can be analysed using differential equations. We will call this the epidemic phase. It is studied in section \ref{EP}.
 
 Finally there is a fading of phase when the epidemic has almost reached it final state. In this phase those still infectious has a small chance of contacting a susceptible individual. The spread will slowly fade off. We will not be much concerned about this phase but it is shortly discussed in section \ref{FP}.

\section{The Branching process phase}\label{BP}

In the start, i.e., before the fact that the population is finite and that the infection causes immunity influences the spread, we can approximate the epidemic process with a branching process. 

There is a linguistic problem. The terminology referring to infections and contacts does not fit well with how branching processes are usually presented. The theory of branching process are closely connected with ideas from demography. It is natural to refer to the events as births and to talk of mothers and offsprings or children (instead of infectors and infected). We will use this terminology in this section.We assume that a newborn child first goes through a latent time during which it can not have any children and then a fertile time during which it gives births to children according to a homogeneous Poisson process, and finally dies. The latent and fertile times correspond to the latent and infectious time in the epidemic model.
 
We will start by studying the properties of a branching process defined by the same sequence of times and the same contacts as the epidemic process.  Then we will discuss the relation between the branching process and the epidemic process.

The branching process is a special case of a so-called Crump-Mode-Jagers-process (i.e. a CMJ-process), see \cite{CM1}, \cite{CM2}, and \cite{Jagers} . This process is can be described in demographic terms where an individual gets offsprings according to a (general) point process during its, possible, random life time. 

Crump-Mode-Jagers-processes  have been studied in great detail and much is known of their properties.
In the following subsection we will  relate some important results  for such processes under the special  assumptions here made. 

When this is done we will use the branching process to construct a related process that have the stochastic properties of the epidemic process. This will make it possible to use the results for branching processes to derive results for the epidemic processes.

\subsection{Results for a branching process}

We will define the branching process as a counting process, $B$,  with the intensity

\begin{equation}\label{intensity2}
\lambda\left[\int\limits_0^t r_{B(s)}(t-s)dB(s) + r_0(t)\right].
\end{equation}
where the functions $r_i$ are defined by (\ref{infect}). The branching process, $B$, differs from the epidemic process, $N$, since it is not influenced by immunity.

Let
\begin{equation}\label{ineqmean}
M(t)= {\rm E}(B(t)).
\end{equation}
The following analysis of asymptotic properties of the process $B$ will repeat some basic, well-known results (see e.g. \cite{Harris}, \cite{Kimmel}, and \cite{Haccou}). Derivations here are based the special structure due to the assumption that the births occur according to a homogeneous Poisson process during the random fertile time. This assumption simplifies derivations of results that are valid in more complex models. 

The most basic results are summarized in two theorems.
Let

\begin{equation}
Z(t) = B(t)e^{-\alpha t}
\end{equation}
where $\alpha$ is the Malthus parameter. The first theorem follows more generally from results in \cite{Feller}. 

\begin{theorem}  
  \begin{equation}\label{meaneqt}
  \lim_{t\to\infty} {\rm E}(Z(t)) \to -\frac{1}{\alpha R_0 L'_g(\alpha)}.
  \end{equation}
  \end{theorem}
 
 \noindent\begin{proof}
 
 We will use properties of a homogeneous  Poisson process.   
The development of the process will depend how the initial mother gives birth to children. The function $r_0(t)$ indicates if the first mother  is fertile at time $t$ or not. 
\begin{equation}
{\rm E}(r_0(t)) = {\rm E}(X)g(t),
\end{equation}
The total time the first mother has been fertile up till time $t$ equals:
\begin{equation}
i(t)=\int\limits_0^t r_0(s)ds =X_0\wedge (t-Y_0)^+.
\end{equation}

We will first condition on $Y_0$ and $X_0 $. With this conditioning the number of children of the initial mother up till time $t$, denoted by, $\tilde R(t)$, is Poisson distributed with mean $\lambda  i(t)$. Due to properties of a Poisson process births occurs at
 $\tilde R(t)$ random times that are independent and uniformly distributed in the interval $[y,y+ i(t)]$. 

Due to the regenerative properties of the process (i.e. each birth starts a new independent stochastically identical process) we have

 \begin{equation}\label{meaneq}
 {\rm E}(B(t)\mid Y_0, X_0,\tilde R(t))=\tilde R(t)\left(1 +\frac{1}{ i(t)}\int\limits_y^{y+ i(t)}M(t-u)du\right).
 \end{equation}
  The last term equals $0$ if $i(t)=0$.
  
  Removing the conditioning on $\tilde R(t)$ we obtain:
 \begin{equation}\label{sst}
 {\rm E}(B(t)\mid Y_0,X_0)=\lambda i(t)+\lambda \int\limits_y^{y+ i(t)}M(t-u)du.
 \end{equation}
 
Since
\begin{equation}
\int\limits_y^{y+ i(t)}M(t-u)du= \int\limits_0^t r_0(s)M(t-s)ds,
\end{equation}
it follows that
\begin{equation}
{\rm E}(B(t)) = M(t) = R_0\left(G(t) + \int\limits_0^{t} g(s)M(t-s)ds\right).
\end{equation}
 Thus the mean of the branching process is only a function of $R_0$ and the generation time distribution $g$.

 If we multiply  equation (\ref{sst}) with $\exp(-ts)$ where $s>\alpha$   we obtain
 
 \begin{equation}\label{meaneq2}
\int\limits_0^{\infty}{\rm E}\left(B(t)e^{-st}\right)dt= \frac{R_0L_g(s)}{s} + R_0L_g(s)\int\limits_0^{\infty}{\rm E}\left(B(t)e^{-st}\right)dt.
\end{equation}
Thus
\begin{equation}
\frac{1-R_0L_g(s)}{s-\alpha}(s-\alpha)\int\limits_0^{\infty}{\rm E}\left(B(t)e^{-st}\right)dt=\frac{R_0L_g(s)}{s}.
\end{equation}
The right hand side of this equation tends to 
$R_0L_g(\alpha)/\alpha = 1/\alpha$ (see \ref{me}) and
\begin{equation}
\frac{1-R_0L_g(s)}{s-\alpha}\to -R_0L'_g(\alpha),
\end{equation}
as $s\to \alpha$. Thus
\begin{equation}
(s-\alpha)\int\limits_0^{\infty}{\rm E}\left(B(t)e^{-(s-\alpha)t}e^{-\alpha t}\right)dt\to  -\frac{1}{\alpha R_0 L'_g(\alpha)}
\end{equation}
as $s\to\alpha$.

A Tauberian theorem says that $r\int\limits_0^{\infty}Q(t)\exp(-rt)\to Q$ as $r\to 0$ if and only if 
$Q(t)\to Q$ as $t\to \infty$, This finally proves the theorem.

 \vskip 0.4 cm
 
The above theorem describes how $M(t)$ grows for large $t$. We will later need an inequality valid for all $t >0$.
The latent times and the finite fertile times are  important in the development of the branching process. If we compare with a process with no latent time and infinite fertile times where births occur with the constant intensity, $\lambda$, it is obvious that this process is stochastically larger than the branching process defined by (\ref{intensity2}). The larger process is  a pure birth process with birth intensity $\lambda$. This implies that
\begin{equation}
M(t)\leq e^{\lambda t} -1.
\end{equation}
Combining this with the asymptotic result from theorem \ref{meaneqt} we obtain:

\begin{corollary}\label{cor}
There exists a constant $K$ such that
\begin{equation}
M(t) \leq Ke^{\alpha t}
\end{equation}
for all $t\geq 0$.
\end{corollary}

 The following theorem gives a relation for the Laplace transform of the limit distribution of $B(t)\exp(-\alpha t)$. The proof again uses the properties of  Poisson processes. Observe the similarity to results by \cite{Harris} for slightly different models. 

\end{proof}
\vskip 1cm
  
 \begin{theorem}\label{inprob}
 \begin{equation}
 L_Z(s)=\lim_{t\to\infty}{\rm E}(e^{-sZ(t)}),
 \end{equation}
 satisfies
\begin{equation}\label{basic}
L_Z(s) =\int\limits_0^{\infty}\int\limits_0^{\infty}\exp\left(-\lambda(x-\int\limits_y^{y+x}L_Z(se^{-\alpha u})du)\right)f_{YX}(y,x)dydx.
\end{equation}
Together with (\ref{meaneq}) this defines $L_Z(s)$ and the limit distribution of $Z(t)$ as $t\to\infty$.
\end{theorem}

 \noindent\begin{proof}
 First we analyze what happens conditional  on the latent time, $Y_0$, and fertile time, $X_0$ for the first mother. The first mother has $\tilde R$ children, where $\tilde R$ is Poisson-distributed with mean $\lambda X_0$. The births occur at times which are independent and uniformly distributed in the interval $[Y_0,X_0+Y_0]$. Thus
 \begin{equation}
 {\rm E}(e^{-sZ(t)}\mid X_0=x,Y_0=y,\tilde R=r) = e^{-sre^{-\alpha t}}\big[\int\limits_y^{x+y}{\rm E}(e^{-sZ(t-u)e^{-\alpha u}})du\big]^r.
 \end{equation}
 Removing the conditioning, first on $\tilde R$, and then on $X_0$ and $Y_0$, and finally letting $t\to\infty$ we derive  the expression (\ref{basic}).
\end{proof}
\vskip 0.5cm
\noindent
From equation (\ref{basic}) we can derive interesting limit results.
E.g. $L_Z(\infty)=q=1-p$ where $p$ satisfies
\begin{equation}\label{short}
1-p=L_X(\lambda p).
\end{equation}
This gives the probability, $q=1-p$, that the process stays finite, see (\ref{small}).
We can write
\begin{equation}\label{division}
L_z(s)= q+pK(s)
\end{equation}
where $K$ is the Laplace transform of the limit of $Z(t)$ given that the process $B(t)$ grows asymptotically large.
Inserting (\ref{division}) in (\ref{basic}) we obtain:

\begin{equation}\label{stor}
q+pK(s) = \int\limits_0^{\infty}\int\limits_0^{\infty}\exp\left(-\lambda p(x-\int\limits_y^{y+x}K(se^{-\alpha u})du)\right)f_{YX}(y,x)dxdy.
\end{equation}

The asymptotic mean of $Z(t)$ as $t\to\infty$ is given by theorem (\ref{meaneq}).
The second moment given that the limit is positive, i.e. the process grows large, can be derived from  the equation:

\begin{eqnarray}\label{second}
\frac{1}{2\alpha}K''(0)\left[2\alpha - \lambda\lbrace L_Y(2\alpha)- L_{X+Y}(2\alpha)\rbrace\right] = \nonumber \\
\frac{p\lambda^2}{\alpha^2}K'(0))^2[ L_Y(2\alpha)-2L_{2Y+X}(\alpha)+L_{X+Y}(2\alpha)].
\end{eqnarray}

Here
\begin{eqnarray}
{\rm E}(Z\mid \hbox{the process grows large}) = K'(0)\\
{\rm E}(Z^2\mid \hbox{the process grows large}) = K''(0)
\end{eqnarray}

It is, in general, difficult to derive an explicit solution of equation (\ref{stor}). However, in the special case that there is no latent time, i.e. $Y\equiv 0$ it is possible. In that case
\begin{equation}
p\lambda=\alpha,
\end{equation}
and we can verify that the Laplace-transform
\begin{equation}
K(s)=\frac{1}{1 + \gamma s}
\end{equation}
satisfies the equation (\ref{basic}). 

The asymptotic expression of the mean of $Z(t)$ given that it is large is asymptotically large is
\begin{equation}
\gamma= -\frac{1}{p\alpha R_0 L'_g(\alpha)}. 
\end{equation}
Thus the limit given that the process grows large is exponential distributed with intensity $1/\gamma$.

For the case there the fertile times are exponentially distributed this is well-known. \cite{KendD} derives an expression for the Laplace transform of the distribution of $Z(t)$, i.e. not only the limit distribution.

Also in cases when there is a positive latent time the limit distribution may be exponential distributed. Later in this paper we will give some examples.

\subsection{Deriving the epidemic process from the branching process}

Two important features differs between the model for epidemic spread and the branching process model:

\begin{itemize}
\item The epidemic  takes place in a finite population,

\item In the epidemic model the growth of the process is slowed down by immunity.
\end{itemize} 
Both these  features have to be considered when relating  the two models.

Starting from an idea from \cite{BD95} we will the construct the epidemic process as follows. Each event (child) in the branching process is marked with a number. The initial mother has the mark $0$.
At birth a child choose, at random, with equal probability, one of the $n$ persons in the population. To the child is then attached a personal number that denotes how many times the chosen person has been previously chosen by older children. Finally the child is marked by a number that is the sum of the personal number of the child and the mark of its mother. 

This mark is $0$ if the child, and all its ancestors, have chosen a person that has not been chosen before. If this is not the case the mark is a number $\geq 1$. The events (births) with mark $0$ will correspond to true infections. If we, in the branching process, delete all births with mark $\geq 1$ the remaining events will describe a process equivalent to the epidemic process. We will denote, in this section, this process with $N_n$, indicating the size of the population. The construction guarantees that a person is infected at most once and that these infections are the only ones that contribute to the spread.

We can now use properties of the branching process to derive properties of an epidemic process. First we observe that at the time when $B(t)=r$ the probability that child $r$ chooses a person already chosen is less than $r/n$. From this we conclude that the probability that the epidemic process differs from the branching process up till the time when $B(t)=r$, smaller than
\begin{equation}
\prod\limits_{i=1}^r (1-i/n) =  \frac{(n-1)!}{(n-r-1)!n^r}
\end{equation}
Applying Stirlings formula we find that this probability tends to $0$ as $n\to\infty$ if $r=n^a$ where $a<1/2$.
Thus the epidemic process has asymptotically the same properties as the branching process till $n^a$ persons have been infected. This is a well known result from \cite{BD95}. The epidemic process and the branching process shares the properties which depends on what happens before the time when $N_n(t)=B(t) \leq n^a$ where $a<1/2$.

However, we will need to use the branching process as an approximation for a longer time. 
To do this we will approximate how many future births are removed if a child has the personal number $\geq 1$.

From theorem \ref{inprob} it follows that $B(t)exp(-\alpha t)$ converges in distribution to a random variable $Z$. For the following argument we will need the stronger result that that there exists a random variable $Z$ such that  $B(t)exp(-\alpha t)  \to Z$ with probability $1$ as $t \to \infty$. This result is not proved in this paper but can be found in several places, e.g. \cite{CM2}.

We will first consider how many  of the $m_n > r_n$ first born children are direct descendants of the $r_n$`th child. It is assumed that $r_n$ are so large that we can consider $B(t)exp(-\alpha t)$ to be almost constant.
The expected number of direct descendants will be denoted by $V_{r_n}^{m_n}$. 

The time it takes for the process $B$ to go from $r_n$ to $m_n$ is close to $\ln(m_n/r_n)/\alpha$. The direct descendants of the $r_n$`th child develops according to the same rules as the branching process.  From corollary \ref{cor} we find that
\begin{equation}\label{vmean}
V_{r_n}^{m_n}\leq K \frac{m_n}{r_n}.
\end{equation}

We have already observed that there with probability tending to $1$  do not exist any child with personal number $\geq 1$ among the $n^a$ first born if $a< 1/2$. The probability that the $r_n$`th child child has personal number $\geq 1$ is less than $r_n/n$. Using the inequality (\ref{vmean}) and including the $r_n$`th child we find that the expected number of children that are not included in the epidemic process due to that the $r_n$`th child has personal number $\geq 1$ is at most $ (V_{r_n}^{m_n}+1)r_n/n$. This gives a crude inequality for the expected difference between $B$ and $N_n$ up till the time where $B(t)=m_n$ , namely
\begin{equation}
B(t_{m_n}) - \rm E(N_n(t_{m_n})) \leq \sum\limits_{v =n^a}^{m_n} \frac{v}{n}\left(K\frac{m_n}{v}+1\right) \leq (K+1/2)\frac{m_n^2}{n}.
\end{equation}

Obviously $B(t)-N_n(t)$ is always non-negative and also increasing in $t$. We can now apply the Markov inequality and draw the following conclusions:

\begin{itemize}
\item
\begin{equation}
\frac{N_n(t)}{B(t)} \to 1
\end{equation}
in probability as $n\to \infty$ for all $t$ such that $B(t)\leq n^b$ where $b<1$.

\item
for any $\epsilon >0$ there exists a $\eta$ such that
\begin{equation}
\frac{N_n(t)}{B(t)} \geq 1-\epsilon
\end{equation}
as $ n\to \infty$ for all $t$ such that $B(t)\leq \eta n$.
\end{itemize}

\section{The epidemic phase}\label{EP}
.
Assume that the epidemic grows large and a non-negligible proportion are infected, then at some finite time $\tau_{\epsilon}$ the number of infected will reach the level $n\epsilon$. If $\epsilon$ is sufficiently small the process $N(t)$ can be well approximated by a branching process up till that time. After that the process has to be studied by other methods.

Using the branching process approximation we find that
\begin{equation}
N(\tau_{\epsilon})exp(-\alpha\tau_{\epsilon})=n\epsilon exp(-\alpha\tau_{\epsilon})\approx Z
\end{equation}
where $Z$ is a random variable. This implies that
\begin{equation}\label{time}
\tau_{\epsilon}=\frac{\ln(n)}{\alpha} + \tilde Z.
\end{equation}
where $ \tilde Z$ is a finite random number.

More exact results (and more rigid analysis) valid in special models can be found in \cite{Bar} and \cite{Svens3}.

 We start by defining the new counting process $\bar N(s)$ that counts the number of infections that take place after $\tau_\epsilon$, i.e., 
$\bar N(s)=N(s+\tau_{\epsilon})-N(\tau_{\epsilon})$. This new counting process has the intensity
\begin{eqnarray}
\bar \lambda(s) &=& \lambda(1-\epsilon-\frac{\bar N(s-)}{n})\int\limits_0^sr_{N(v+\tau_{\epsilon})}(s-v)d\bar N(v)  \\
&+&  \lambda(1-\epsilon-\frac{\bar N(s-)}{n})\left[\int\limits_0^{\tau_{\epsilon}}
r_{N(u)}(s+\tau_{\epsilon}-u)dN(u) + r_0(s+\tau_{\epsilon})\right].\nonumber
\end{eqnarray} 
The first right-hand term gives the intensities of infections caused by those infected after $\tau_{\epsilon}$ and the second term the intensity caused by those infected before that time but occurring after $\tau_{\epsilon}$. It is thus  necessary to investigate the effect of the remaining infectivity spread by those infected before $\tau_{\epsilon}$ after that time.

\subsection{Remaining infectivity}

We will now consider, using an heuristic argument, how much infectivity has been spread in the population at time $\tau_{\epsilon}$, when  
$\epsilon n$ individuals have been infected and how much infectivity is still remaining to be spread after time $\tau_{\epsilon}$ by those infected before $\tau_{\epsilon}$.

The infections occurs according to a Poisson process. Thus we can assume that  it requires, in mean, a total of infectiousness  $\epsilon n$ to produce $\epsilon n$ infections when we can disregard effects of immunity. The first $\epsilon n$ infected can, in mean, generate the infectiousness $R_0\epsilon n$. Thus, at the time $\epsilon n$ individuals has been infected, if that happens, there still remains, in mean, $(R_0-1)\epsilon n$ infectiousness to be spread from those already infected. We will  consider how this infectivity is distributed in time after $\tau_{\epsilon}$.
This, of course, depends on when those infected before $\tau_{\epsilon}$ are infected.

We will try to obtain a useful expression of
\begin{equation}
\frac{\lambda}{n}\left[\int\limits_0^{\tau_{\epsilon}}
r_{N(u)}(s+\tau_{\epsilon}-u)dN(u) + r_0(s+\tau_{\epsilon})\right].
\end{equation}
The expected value of this expression is
\begin{equation}\label{rest}
R_0\frac{e^{\alpha\tau_{\epsilon}}}{n}\left[\int\limits_0^{\tau_{\epsilon}}
g(s+\tau_{\epsilon}-u)e^{-\alpha(\tau_{\epsilon}-u)}{\rm E}\left(e^{-\alpha u}dN(u)\right)\right]+\frac{\lambda}{n} r_0(s+\tau_{\epsilon}).
\end{equation}
Now
\begin{equation}
\frac{e^{\alpha\tau_{\epsilon}}}{n}=\frac{\epsilon}{N(\tau_{\epsilon})e^{-\alpha\tau_{\epsilon}}}.
\end{equation}

Another approximation yields
\begin{equation}
{\rm E}(e^{-\alpha u}dN(u))\approx {\rm E}(\alpha N(u)e^{-\alpha u}du).
\end{equation}

The remaining infectivity from the initial infector will disappear as $t\to \infty$. Now ${\rm  E}(N(t)e^{-\alpha t})$ approaches a constant  conditional on that the process grows asymptotically large.  Inserting these approximations in (\ref{rest}) we find, that the infectivity remaining from those infected before time $\tau_{\epsilon}$ spread out in time is $R_0\epsilon nRem(s)$ where
\begin{equation}
Rem(s) \approx \left[\int\limits_0^{\tau_{\epsilon}}
g(s+\tau_{\epsilon}-u)e^{-\alpha(\tau_{\epsilon}-u)}\alpha du\right].
\end{equation}

Since $\tau_{\epsilon}\to \infty$ as $n\to\infty$ according to (\ref{time}) we can use the approximation
\begin{equation}
Rem(s) =  \alpha\int\limits_0^{\infty}g(s+t)e^{-\alpha t}dt
\end{equation}
if $n$ is large.

Integrating the second right-hand term of this equation we get
\begin{equation}
R_0\int\limits_0^{\infty}Rem(t)dt = R_0\alpha\int\limits_0^{\infty} (1-G(t))e^{-\alpha t}dt = R_0-1
\end{equation}
which corresponds to the observation of the amount of the remaining infectiousness made at the beginning of the section.
Also observe that
\begin{equation}
R_0Rem(0)=\alpha.
\end{equation}
This has to be the case, since the process is assumed to have the Malthus parameter $\alpha$.

\subsubsection{Some simple examples}

We will illustrate the calculations necessary by considering different sets of assumptions of latent times, infectious times and infectivity. 

In the first two examples we assume that there is no latent time, i.e. $Y\equiv 0$. In this situation we can derive the limit distribution of $Z(t)$ from the results above. In the third example there are exponentially distributed latent times and  exponentially distributed infectious times. In that case the limit distribution may or may not be exponentially distributed, depending on the parameter values.
\vskip 1 cm
\noindent{\it Exponentially distributed infectious time}

The basic assumptions are that $Y\equiv 0$, and $X$ exponentially distributed with intensity $\beta$ and the infectivity is $\lambda$. This gives:

\begin{equation}
R_0 =\frac{\lambda}{\beta}.
\end{equation}

\begin{equation}
g(t) = \beta e^{-\beta t}.
\end{equation}

\begin{equation}
 L_g(s) = \frac{\beta}{\beta+s}.
\end{equation}

\begin{equation}
T =\frac{1}{\beta}.
\end{equation}

\begin{equation}
\alpha = \lambda - \beta.
\end{equation}

\begin{equation}
p=\frac{\alpha}{\lambda}.
\end{equation}

Given that the epidemic grows large $Z(t)$ is asymptotically exponentially distributed with intensity $1/\gamma$ where
\begin{equation}
\gamma = (\frac{\lambda}{\alpha})^2.
\end{equation}

\begin{equation}
Rem(s) =\frac{\alpha\beta}{\lambda} e^{-\beta s}.
\end{equation}
It should be observed thats
\begin{equation}
R_0Rem(s)\equiv \alpha g(s)/\beta.
\end{equation}
 This implies that at time $\tau_{\epsilon}$ there are $n\epsilon\alpha/\beta = (R_0-1)\epsilon n$ infectious individuals each being as infective as a newly infected individual. This is, of course, due to the ``lack of memory'' property that characterizes the exponential distribution.
 \vskip 1cm

\noindent{\it Constant infectious time}
 
 The basic assumptions are that $Y\equiv 0$, and $X\equiv k$ where $k$ is a constant, i.e., there is a constant, non-random, infectious time. Such a process is sometimes referred to as a (continuous time) Reed-Frost process. The infectivity is $\lambda$. This gives:
 
\begin{equation}
R_0 =\lambda k.
\end{equation}

\begin{equation}
g(t) = \frac{I(t<k)}{k}.
\end{equation}

\begin{equation}
 L_g(s) = \frac{1-e^{-sk}}{sk}.
\end{equation}

\begin{equation}
T =\frac{k}{2}.
\end{equation}

The Malthus parameter, $\alpha$, solves the equation 
\begin{equation}
 \lambda(1-e^{-\alpha k})=\alpha.
\end{equation}

\begin{equation}
p=\frac{\alpha}{\lambda}.
\end{equation}

Given that the epidemic grows large $Z(t)$ is asymptotically exponentially distributed with intensity $1/\gamma$ where
\begin{equation}
\gamma = \frac{\lambda}{\alpha(1-\lambda ke^{-\alpha k})}=\frac{\lambda}{\alpha(1-R_0+\alpha k)}.
\end{equation}

\begin{equation}
Rem(s) =\frac{\alpha}{k} (1-  e^{-\alpha (k-s)^+}).
 \end{equation}
 
\vskip 1cm
\noindent{\it Exponentially distributed latent and infectious times}
 
The basic assumptions are that $Y$ is exponentially distributed with intensity $\delta$, and  $X$ exponentially distributed with intensity $\beta$. $X$ and $Y$ are assumed to be independent. The infectivity is $\lambda$. This gives:

\begin{equation}
R_0 =\frac{\lambda}{\beta}.
\end{equation}

\begin{equation}
g(t) = \frac{\delta\beta}{\delta-\beta}(e^{-\beta t}-e^{-\delta t}).
\end{equation}
If $\beta=\delta$ then
\begin{equation}
g(t)=\beta^2te^{-\beta t}.
\end{equation}

\begin{equation}
 L_g(s) = \frac{\delta\beta}{(\delta +s)(\beta + s)}.
\end{equation}

\begin{equation}
T =\frac{1}{\delta} + \frac{1}{\beta}.
\end{equation}

The Malthus parameter, $\alpha$, solves the equation 
 \begin{equation}
  \alpha^2+(\delta+\beta)\alpha -\delta(\lambda-\beta)=0.
\end{equation}

\begin{equation}
p=\frac{\lambda - \beta}{\lambda}.
\end{equation}

Given that the epidemic grows large $Z(t)$ has asymptotic mean $1/\gamma$ where
\begin{equation}
\gamma = \frac{(\delta+\alpha)^2(\beta+\alpha)^2}{\delta\alpha(\delta+\beta+2\alpha)(\lambda-\beta)} .
\end{equation}

\begin{equation}
Rem(s) =\frac{\beta\delta\alpha}{\delta-\beta}\left( \frac{ e^{-\beta s} }{\beta+\alpha}-\frac{ e^{-\delta s}}{\delta+\alpha}\right).
 \end{equation}
 If $\beta=\delta$ then
\begin{equation}
Rem(s) = \beta^2e^{-\beta s}\left(\frac{1}{(\beta+\alpha)^2}+\frac{s}{\beta+\alpha}\right).
\end{equation}
 
 \subsection{Illustrated examples}
 
\begin{figure}
\centering
\scalebox{.5}{\includegraphics{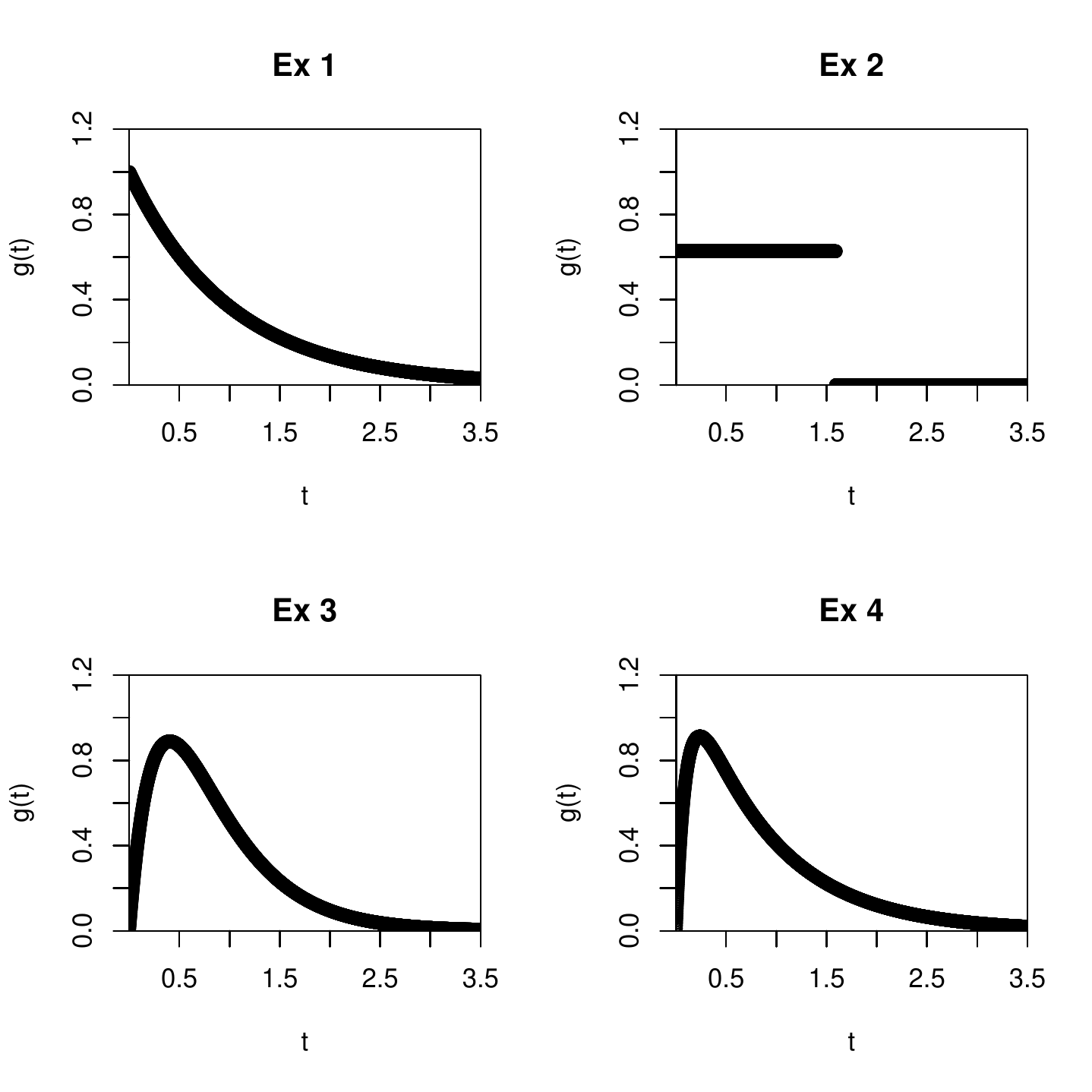}} 
\caption{$g(t)$ for example 1-4}\label{gen}
\end{figure}

\begin{figure}\centering
\scalebox{.5}{\includegraphics{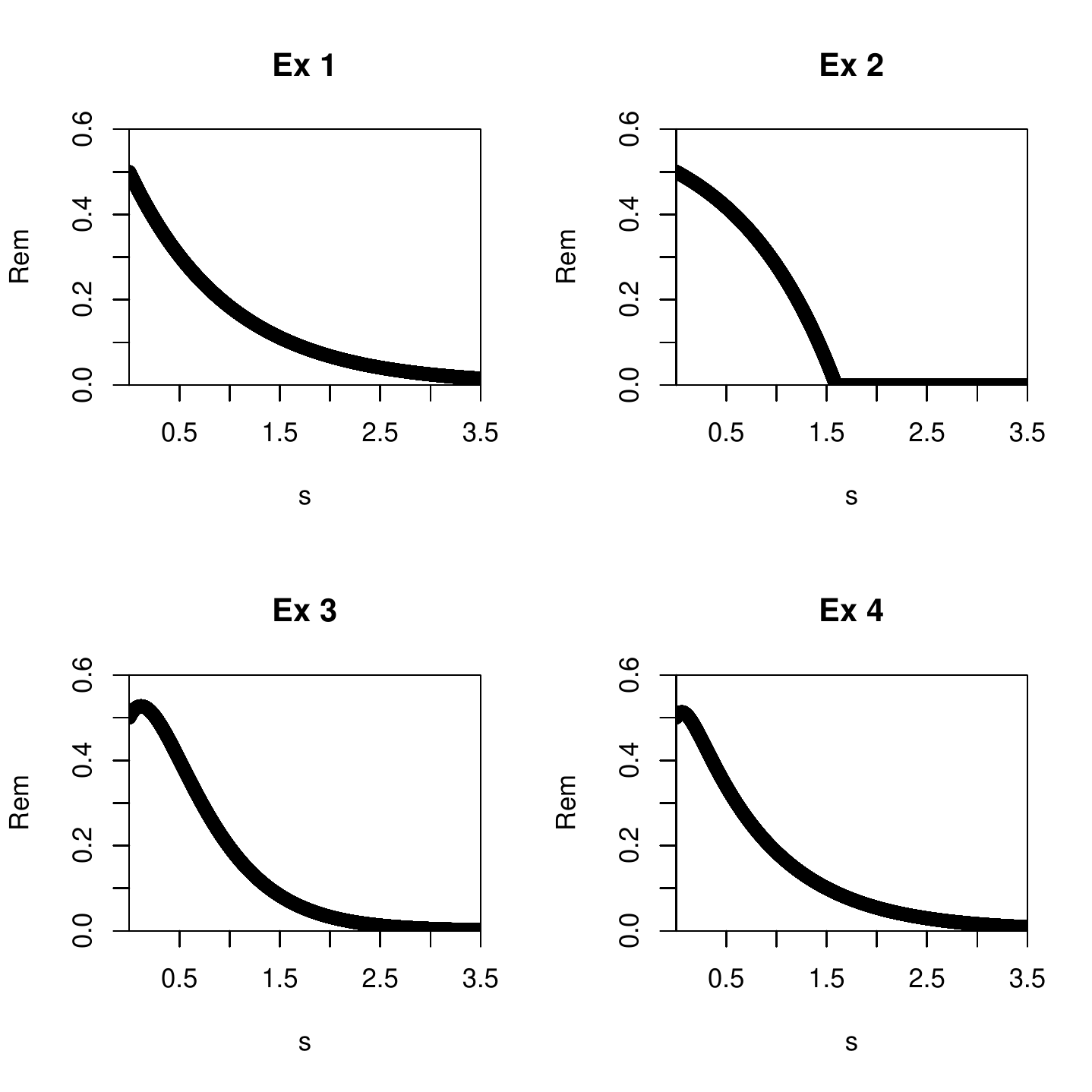}}
\caption{$Rem$ from example 1-4}\label{rem}
\end{figure}

To illustrate the models exemplified above we have chosen parameters values that give the same basic reproduction number and the same Malthus parameter. We have here aimed at the values:
 $$R_0 = 2,$$
 and
 $$\alpha = 1,$$
 With this basic reproduction number the final size $\pi=0.797$ (see \ref{final}).
 To obtain these values we will have to choose different parameter values. We will consider four examples.
 \vskip 0.3cm
 \noindent{\bf Example 1:}
In the model with no latent time and exponential distributed infectious times we choose $\lambda=2$ and $\beta = 1$. 
With these parameter values the probability for a large epidemic is   $p=0.5$. The mean generation time is $T=1$ and the limit distribution of $Z(t)$ is exponential with mean $4$. Furthermore
\begin{equation}
Rem(s) = \frac{e^{-s}}{2}.
\end{equation}
 \vskip 0.3cm
 \noindent{\bf Example 2}:
 In the model with no latent time and constant infectious time we choose
 $\lambda=1.255,$
 and
 $k=1.593.$  The probability for a large epidemic is $\pi=0.797$ and the mean generation time is $T=0.797$. The limit distribution of $Z(t)$ is exponential with mean $2.12$. Also
 \begin{equation}
 Rem(s)=\frac{1-e^{-(1.593-s)^+}}{1.593}
 \end{equation}
 \vskip 0.3cm
\noindent {\bf Example 3}:
 In the model with exponential latent and infectious times we choose first $\beta =2$, $\lambda=4$,  and $\delta=3$. The probability for a large epidemic equals $p=0.5$. The mean generation time is $=5/6\approx 0.833$. The limit distribution of $Z(t)$ is  exponential with mean $24/7\approx 3.43$ and
 \begin{equation}
Rem(s) =2e^{-2s}-1.5e^{-3s}.
\end{equation}
\vskip 0.3cm
\noindent{\bf Example 4}:
  In the model with exponential latent and infectious times we choose first $\beta =10$, $\lambda=20$,  and $\delta=11/9$. The probability for a large epidemic equals $p=0.5$. The mean generation time is $T=0.821$. The limit distribution of $Z(t)$ is in this case not  exponential. It has asymptotic mean $3.698$ and standard deviation $5.39$. This is implies that the limit distribution of $Z(t)$ is not exponentially distributed.
 \begin{equation}
Rem(s) =\frac{110}{79}\left(\frac{9e^{-11s/9}}{20}-\frac{e^{-10s}}{11}\right)
\end{equation}
  \vskip 0.5cm
  \noindent
  In figure \ref{gen} the generation time densities for the four examples are illustrated and in figure \ref{rem} the functions $Rem(s)$ .

\subsection{Differential equation approximation}\label{deter}

Now let
\begin{equation}
x(t)=\frac{\bar N(t)}{n}.
\end{equation}
If we take expectations we find that if $n$ is large
\begin{equation}
\frac{x'(t)}{1-(x(t)+\epsilon)}=R_0\int\limits_0^tg(t-v)dx(v)+ R_0\alpha\epsilon\int\limits_0^{\infty}g(s+t)e^{-\alpha t}dt.
\end{equation}

If we solve this differential equation and let $\epsilon\to 0$ we have an expression for the deterministic trajectory for the progress of the epidemic in the nearly deterministic phase provided the population is large or asymptotically as $n\to\infty$.

\begin{figure}\centering
\scalebox{.5}{\includegraphics{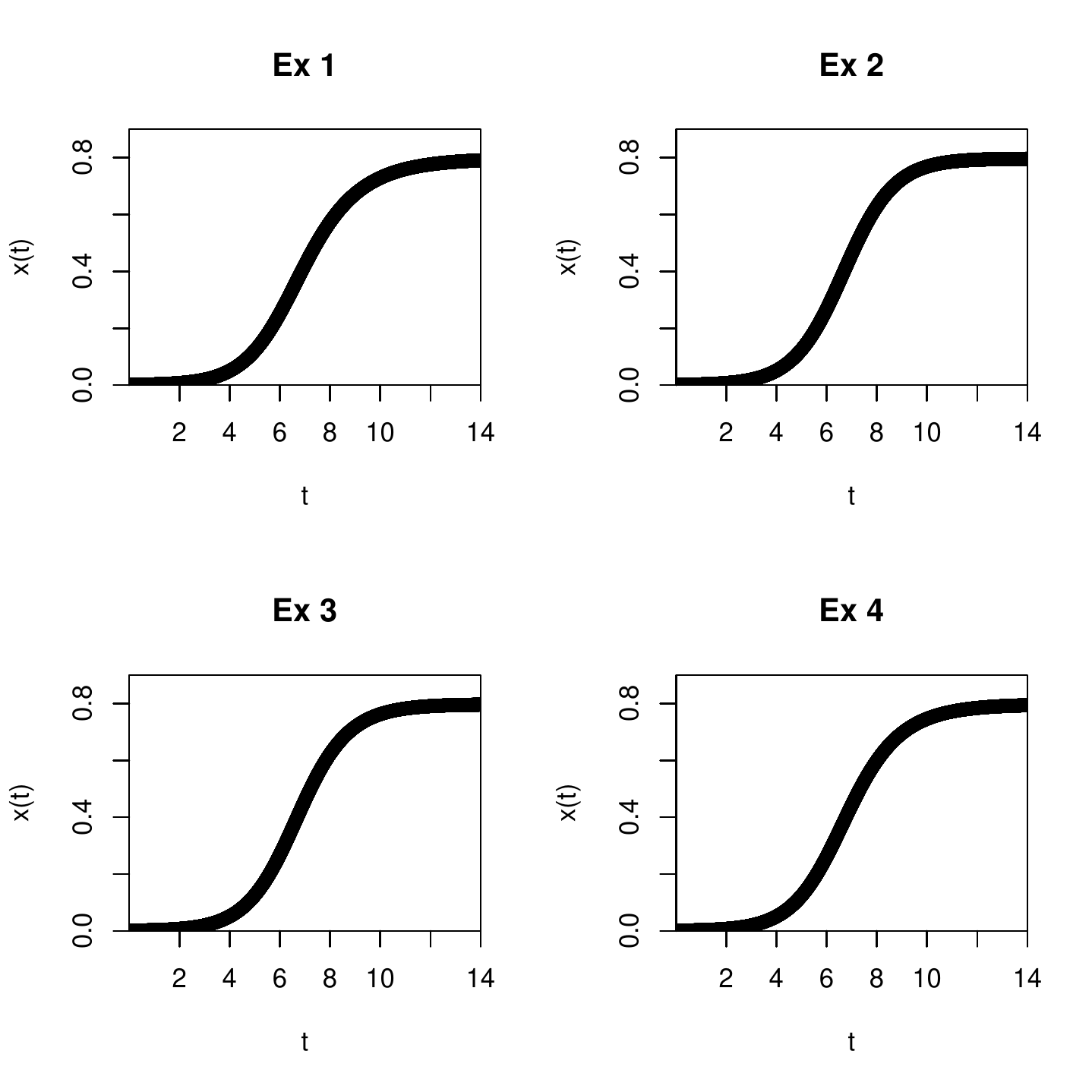}} \caption{Trajectories for epidemics as described by example 1--4}\label{comp}
\end{figure}

In figure \ref{comp} trajectories of the epidemics for the four examples, given that they grow large,  are illustrated. We have chosen values of $\epsilon$ that are proportional to $\gamma$. It is seen that the appearances of the epidemic curves are very similar. This is, of course, due to the fact that we have chosen parameter values so that the final sizes, i.e. $\pi=0.797$ are the same in the four examples and that the initial exponential growth rates, i.e. the Malthus parameters, also are the same.

\section{The Fading off phase}\label{FP}

In the final phase most contacts taken by an infectious individual will be with an immune individual and will not result in further spread of the epidemic. This implies that the process will be essentially random. However we know from (\ref{final}) that finally the epidemic, if it grows large, will end up with the proportion, $\pi$, members infected. 

The fading-off phase will last from the time, $t=\tau_{\pi-\epsilon}$, when $N(t)=(\pi-\epsilon)n$, until it stops. Here $\epsilon$ can be chosen arbitrarily small if $n\to\infty$. 

In the start the epidemic behaves as a branching process, which is a birth-and-death process with, if $R_0>1$, a strong bias to births (i.e. new infections). The final phase the process behave like a birth-and-death process with more deaths (i.e. individuals becoming non-infectious and immune) than births. Of course, it is of interest to study the final phase of the epidemic, but it will not be done in any detail  here.

 \section{Competing epidemics}\label{CE}
 
The results presented above can be used  to study competing epidemics. We will consider what happens if two strains of  an infectious agent  enters into a population simultaneously. It is assumed that an individual can only be infected by one of the strains. An infected individual is immune to further infections, by either strain.
 
 The following discussion is heuristic. In the special case where both strains has no latent time and exponentially distributed infectious times the problem has been studied in detail by \cite{svetom}. The differential equation approximation used in that paper relies on results from \cite{KendW}.
 The present discussion is similar, but less rigid.

First observe that initially there is only a small probability that an infected person tries to infect an immune person with the strain he is carrying. As long as this is the case the spread of the two strains is well described by two independent branching processes. 
After the initial phase the branching process approximation fails to apply and the processes develop deterministically until a fading off phase is reached. This final phase will not be important when the final proportions of infected by the two strains are considered.
 
In the following we will distinguish the two strains by a sub index. Let, e.g., $R_{0i}$ their respective basic reproduction numbers, and
let $p_i$ be the probability that strain $i$ will grow large if it was the only strain entering the population. In that case it will reach the final size $\pi_i$ which is the positive solution of 
 \begin{equation}\label{final1}
 -\ln(1-\pi_i) = R_{0i}\pi_i,
 \end{equation}
 see equation \ref{final}.

 It is of course of interest to find out the final sizes of the epidemic spread of the two strains. Let $\tilde \pi_i$ be the final proportion of the population infected by strain $i$.  If we population is large then the equation
\begin{equation}\label{two}
-\ln(1-\tilde\pi_1-\tilde\pi_2)=R_{01}\tilde\pi_1 + R_{02}\tilde\pi_2
\end{equation}
has to be satisfied.
One possible solution is $(\tilde \pi_1,\tilde \pi_2)= (0,0)$. This implies that none of the strains succeed in infecting a non-negliable proportion of the population. This will happen with probability $(1-p_1)(1-p_2)$. 

The probability that at least one of the strains causes an epidemic is thus $p_1+p_2-p_1p_2$.

 If $R_{01}= R_{02}$ equation (\ref{two}) has a unique positive solution $\tilde \pi =\tilde\pi_1 +\tilde \pi_2$. We will need further analysis to decide how the final size is divided between the two strains.
 
If $R_{01}\neq R_{02}$ the equation defines a curve of possible  values for the total size $\tilde\pi_1 + \tilde\pi_2$. Also in this case it remains to find out the contribution of each strain.
 
 When considering the competition between two strains we have to
 take several properties of the spread potential into account. The Malthus parameter decides how fast the spread is initially is of fundamental importance.
 
\subsection{$\alpha_1\neq\alpha_2$}\label{different}

It is clear that the fastest spreading strain, i.e. the one with the largest Malthus parameter, will have an advantage. In fact, if the Malthus parameters differs the epidemic of the faster strain will have time to reach its final stage before the other strain can compete. However, there is a possibility that the slower strain still  causes an epidemic outbreak provided it is strong enough to spread among the still susceptible when the faster strain has faded away. 

Assume that $\alpha_1 >\alpha_2$.  If the first strain infects a positive proportion of the population it will reach the level $\epsilon n$ at approximately time $\tau_{\epsilon}\sim\ln(n)/\alpha_1$. At this time the second strain, will if it has not died out early, have infected $\sim n^{\alpha_2/\alpha_1}$. Observe that $\alpha_2/\alpha_1<1$. It will take a finite time after $\tau_{\epsilon}$ for the first strain to almost reach its final size and enter into its fading-off phase (see \cite{Bar} and \cite{Svens3}). At this time still $\sim n^{\alpha_2/\alpha_1}$ individuals are infected by the second strain. However, only the proportion $1-\pi_1$ of the population is susceptible to infection. In order that the second infection can be able to spread further its ``active'' basic reproduction number, $R_{02}(1-\pi_1)$ has to exceed $1$. If it does it will reach a non-negative proportion of the population, otherwise it will not be able to infect a positive proportion of the population.

To summarize the possible solutions of \ref{two}. Strain $1$ causes a large outbreak with probability $p_1$. Let $\pi_1$ be the possible solution of equation \ref{final1}.
If 
\begin{equation}\label{weak}
R_{02}\leq \frac{1}{1-\pi_1}
\end{equation}
then
\begin{equation}
\tilde\pi_2=0.
\end{equation}
then  strain $2$ can not, cause a large epidemic and equation \ref{two} has the solution $(\pi_1,0)$.

With probability $(1-p_1)p_2$  strain $2$ causes a large outbreak but not strain $1$. Equation \ref{two} has then the solution 
$(0,\pi_2)$.

With probability $(1-p_1)(1-p_2)$ there will be no outbreak.

It remains to consider the situation when $R_{02}(1-\pi_1)>1$. Even if strain $1$ causes a large epidemic  strain $2$ can also cause a mayor outbreak. Its final size is then the the positive solution of
\begin{equation}\label{twoep}
-\ln(1-\frac{\tilde\pi_2}{1-\pi_1})=R_{02}\tilde\pi_2. 
\end{equation}

Thus four possible outcomes are possible. With probability $(1-p_1)(1-p_2)$ the final sizes will be $(0,0)$. With probability $p_1(1- p_2)$ they will be $(\pi_1,0)$, with probability $(1-p_1)p_2$ it will be $(0,\pi_2)$ and with probability $p_1  p_2)$ it will be $(\pi_1,\tilde \pi_2)$.

\subsection{$\alpha_1 =\alpha_2$}\label{same}
 
 We will now discuss what may happen if the two strains have the same Malthus parameter. The result of the competition will then depend on the possible different strengths, i.e. the basic reproduction numbers and the generation time densities.
 
 Now assume that we consider two strains introduced in the population at the same time. At some time, $ t_0$, they have together infected, and immunized, the proportion $\epsilon$ of the population. We assume that $\epsilon=\epsilon_1 + \epsilon_2$, where $\epsilon_i > 0$ is the proportion infected by strain $i$. If $x_i(t)$ is the proportion  infected by strain $i$ at time $t_0+t$ and $x(t)=X_1(t)+x_2(t)$. 
 
 This gives the differential equations:
 
 \begin{eqnarray}\label{difeq}
 \frac{x'_1(t)}{1-(x(t)+\epsilon)}=R_{01}\int\limits_0^tg_1(t-v)dx_1(v)+ R_{01}\alpha\epsilon_1 Rem_1(t)\\
 \frac{x'_2(t)}{1-(x(t)+\epsilon)}=R_{02}\int\limits_0^tg_1(t-v)dx_2(v)+ R_{02}\alpha\epsilon_2 Rem_2(t).
 \end{eqnarray}
 
By solving this system  differential equations we can find out how the infections are distributed over the two strains. This of course depends on the relation between $\epsilon_1$ and $\epsilon_2$, of the generation time densities $g_1$ and $g_2$, and of the basic reproduction numbers $R_{01}$ and $R_{02}$.

\subsubsection{Simulated examples}
 
 If both strains causes large outbreaks, there is a time $\tau_{\epsilon}$ when the proportion $\epsilon$ of the population has been infected, i.e.
 \begin{equation}
 \frac{N_1(\tau_{\epsilon})+N_2(\tau_{\epsilon})}{n}=\epsilon.
 \end{equation}
 At that time the proportion
 \begin{equation}
 Q=\frac{N_1(\tau_{\epsilon})}{N_1(\tau_{\epsilon})+N_2(\tau_{\epsilon})},
 \end{equation}
 has been infected by strain $i$. Since $\alpha_1=\alpha_2$, $Q$ will, asymptotically, be a random variable with the asymptotic distribution
 \begin{equation}\label{qdef}
 Q=\frac{Z_1}{Z_1+Z_2}
 \end{equation}
 conditional on that $Z_1>0$ and $Z_2>0$. Here $Z_1$ and $ Z_2$ are independent since the two strains do not interact in the start of the processes.
 
 In three of the examples above (and in a fifth considered later) the conditional distributions are exponential. If their means are $\gamma_1$ and $\gamma_2$ then the distribution function of $Q$ is
 \begin{equation}
\frac{q/\gamma_1}{(1/\gamma_1-1/\gamma_2)q+1/\gamma_2}.
\end{equation}
 
In example 4 the conditional limit distribution is not exponentially distributed and we have no explicit expression for the distribution. If a strain with these characteristics is involved in the competition it is possible to simulate branching processes in order to approximate the distribution of $Q$.
\vskip 0.4cm
\noindent
{\it Equal basic reproduction numbers}

 We will first consider  examples where the strains have the same basic reproduction number, i.e. $R_{01}= R_{02}$. If both strains causes large outbreaks the outcome of the competition will  depend on remaining differences in generation functions and on the start of the spread approximated by the branching processes. 

\begin{figure} \centering
\scalebox{.5}{\includegraphics{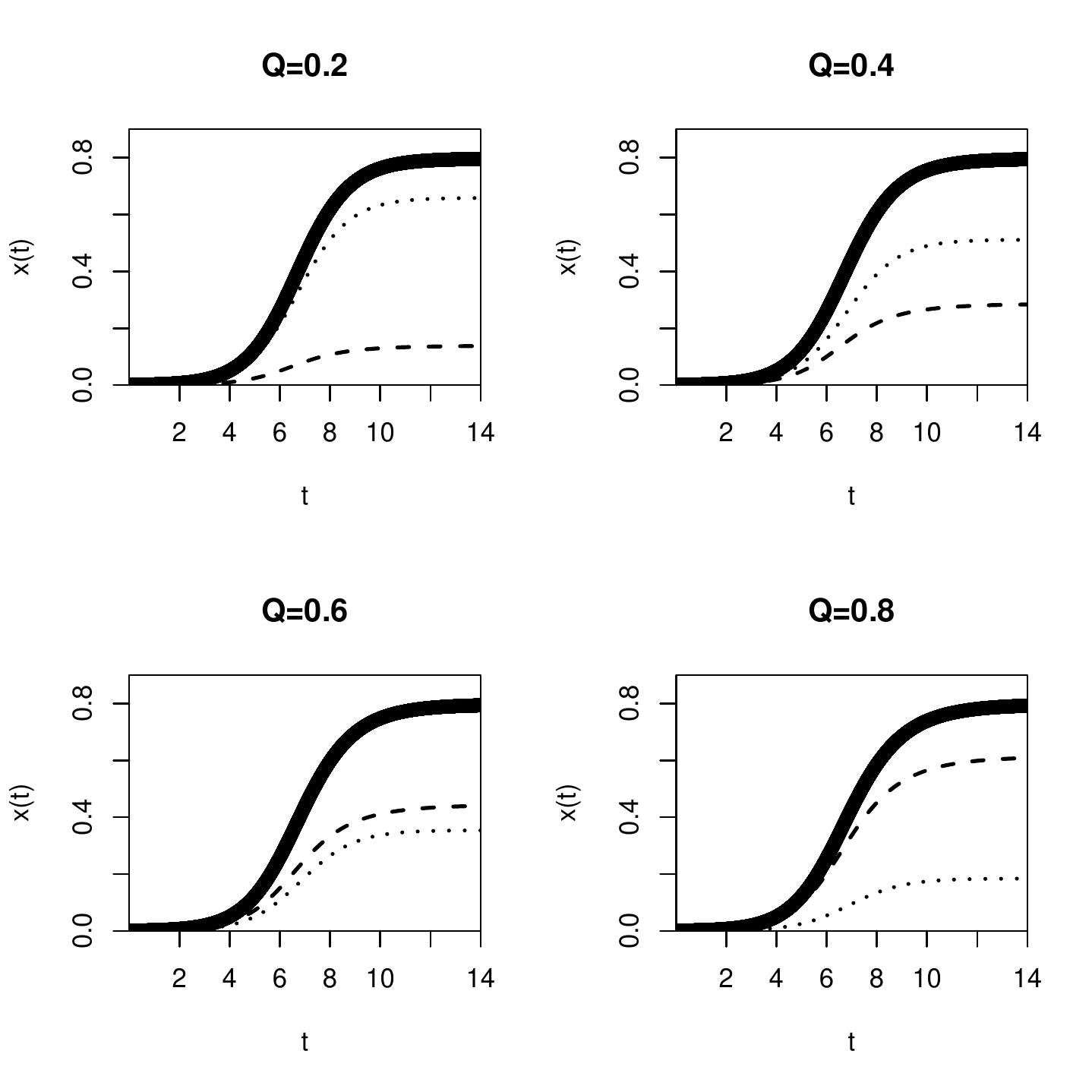}} \caption{The proportion infected in time, when infectiousness of the strain 1 is as in example 1 and of strain 2 as in example 2. The thick line is the total proportion of infected, the dashed line the proportion infected by strain 1 and the dotted line the proportion infected by strain 2.}\label{comp12} 
\end{figure}

In figure \ref{comp12} we illustrates the competition between two strains. One strain spreads according to the assumptions of model 1 and the other as in model 2.  The figure illustrate what happens for different values of the random $Q$. The outcomes illustrated in the figures are conditional to that both strains present causes large outbreaks. It is obvious that it is extremely random which of the strains will dominate in the competition. The relative outcome is essentially decided of what happens early in the two epidemic processes.

\begin{figure} \centering
\scalebox{.4}{\includegraphics{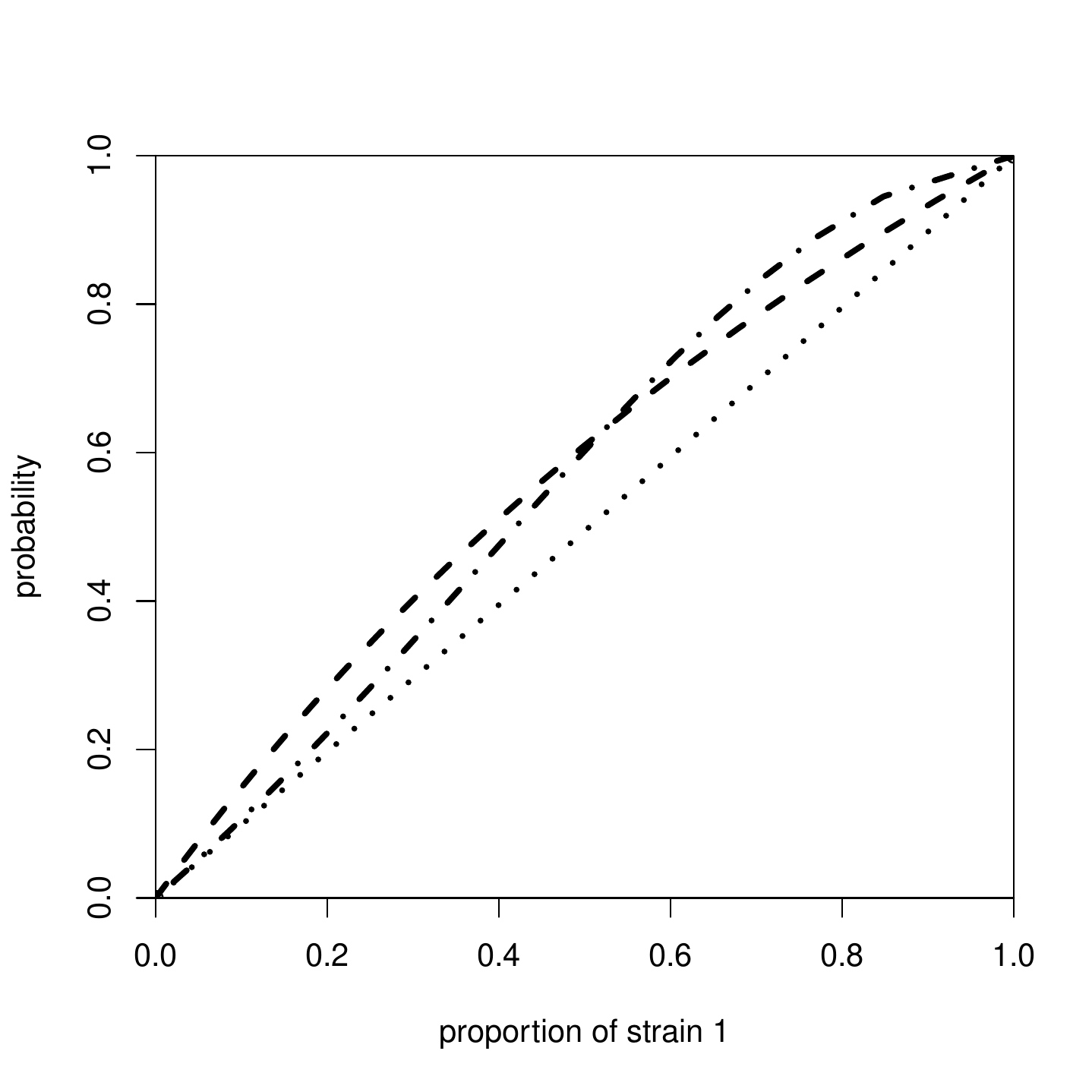}} \caption{Distribution of final proportion of those infected that are infected of strain 1, with infectiousness as in example 1, when the competing strain is as in example 2 (dashed line), example 3 (dotted line) or example 4 (dotdashed line). The distributions are conditional on that both competing strains cause large outbreaks.}\label{equal}
\end{figure}

By solving the differential equations (\ref{difeq}) we can calculate a distribution of the proportion of infected individuals which are infected by strain $1$. This distribution is illustrated in figure \ref{equal}. The distribution is conditional to that both strains causes a large outbreak.

\noindent
{\it Different basic reproduction numbers}

As pointed out above, in the case that the two competing strains have different basic reproduction numbers the proportion of infected by either strain will not be a constant but may take any value on a curve defined by equation (\ref{twoep}). Where on this curve the final state of the epidemics will end is decided by how the strains spreads early in the epidemic, which are well approximated by two independent branching processes.

This situation is discussed in detail by \cite{svetom} for the case where the both strains spreads without latent time and exponential distributed infectious times. We refer to that discussion. As an example we will here illustrate what may happen by an example. One of the strains will behave as in example 3 above and the other as
 \vskip 0.3cm
\noindent {\bf Example 5}:
 In the model with exponential latent and infectious times we choose first $\beta =4$, $\lambda=6$,  and $\delta=5$. The probability for a large epidemic equals $p=1/3$. The mean generation time is $T=0.45$. The limit distribution of $Z(t)$ is with these parameter values  exponential with mean $90/11\approx 8.18$ and
 \begin{equation}
Rem(s) =4e^{-4s}-10e^{-5s}/3.
\end{equation}

\begin{figure} \centering
\scalebox{.5}{\includegraphics{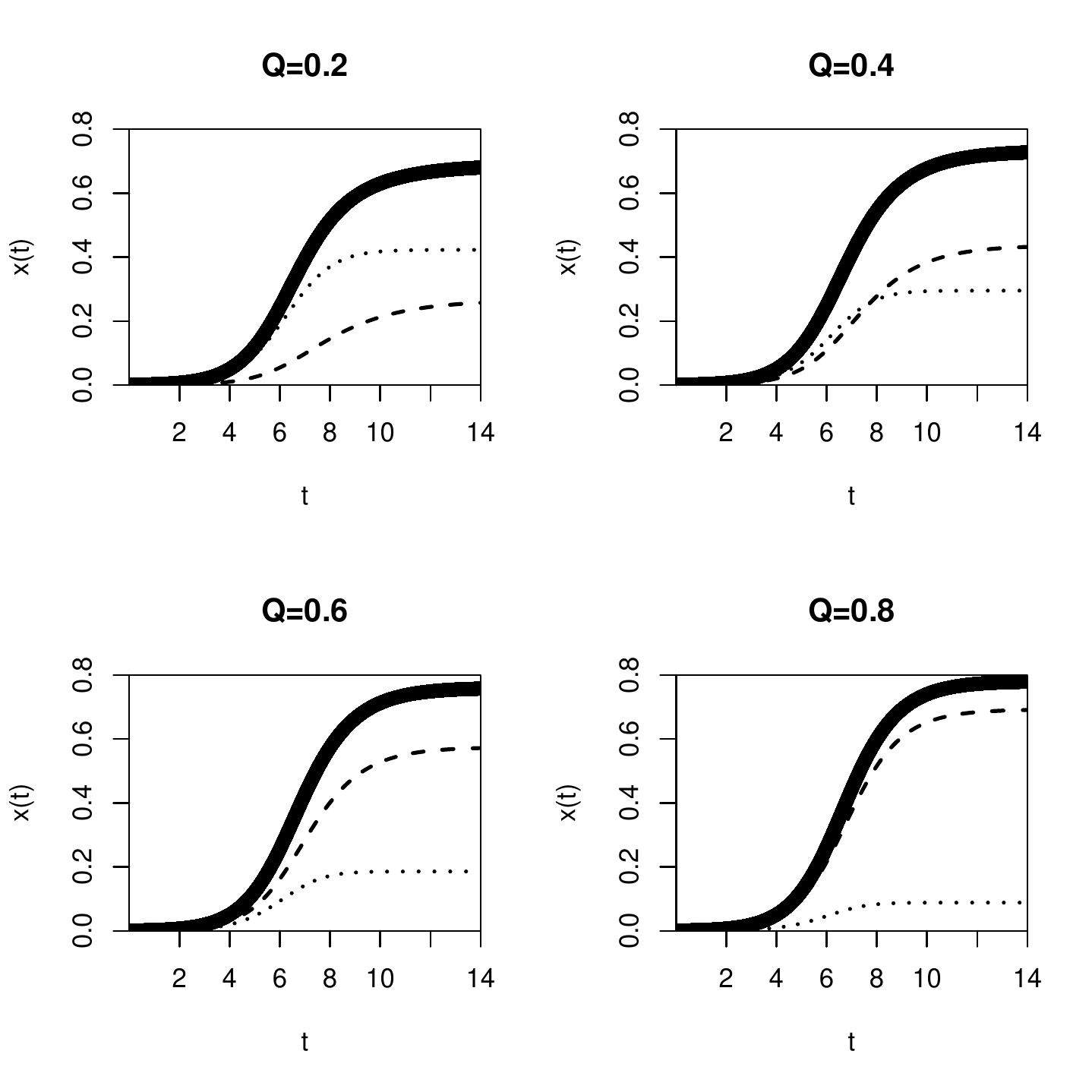}} \caption{The proportion infected in time, when infectiousness of the strain 1 is as in example 3 and of strain 2 as in example 5. The thick line is the total proportion of infected, the dashed line the proportion infected by strain 1 and the dotted line the proportion infected by strain 2.}\label{compete35}
\end{figure}

Figure \ref{compete35} illustrates how the proportion of strain 1 (as in example 3) and strain 2 (as in example 5) develops during the epidemic for different values of $Q$
\begin{equation}
Q=\frac{N_1(\tau_{\epsilon})}{N_1(\tau_{\epsilon})+N_2(\tau_{\epsilon})}.
\end{equation}

\begin{figure} \centering
\scalebox{.4}{\includegraphics{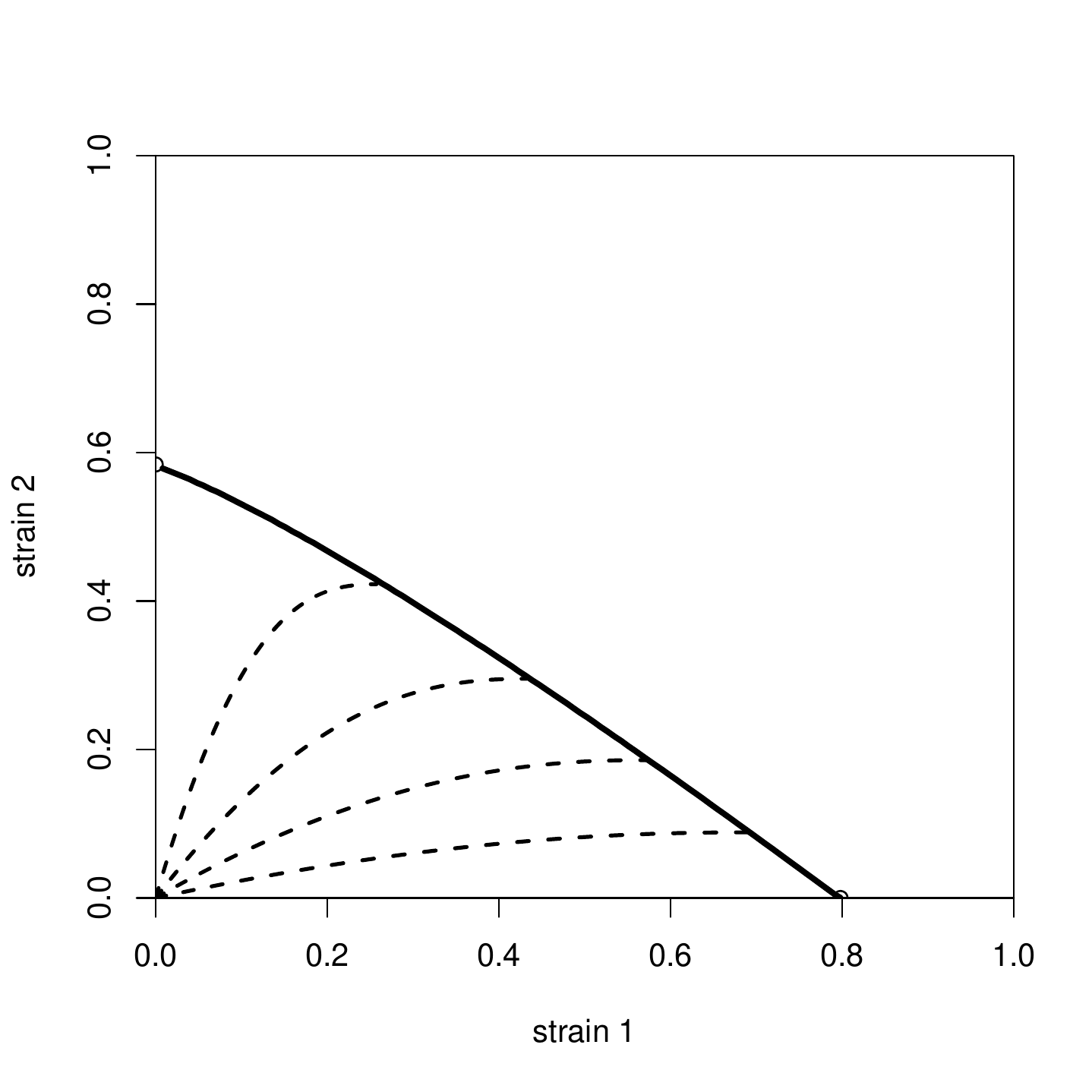}} \caption{Progress of proportion infected of strain 1, with infectiousness as in example 3, and of strain 1, with infectiousness as in example 5. The development are conditional on that both strains cause large outbreaks.}\label{strain35} 
\end{figure}

\begin{figure} \centering
\scalebox{.4}{\includegraphics{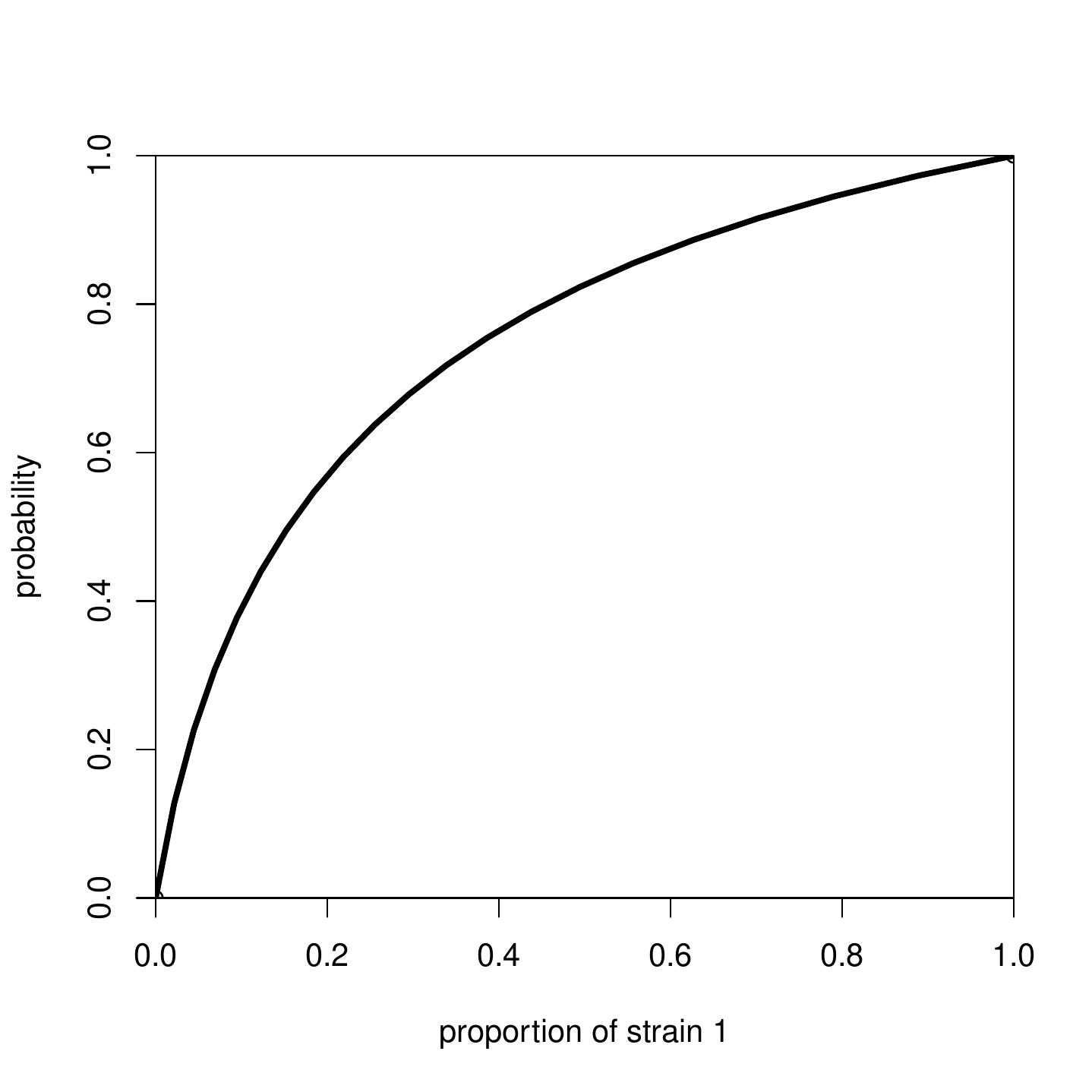}} \caption{Distribution of final proportion of those infected by  a strain  with infectiousness as in example 3, when the competing strain is as in example 5. The distribution is conditional on that both competing strains have large outbreaks.}\label{com35}
\end{figure}

For convenience we have chosen examples 3 and 5 so that branching process limit of $Z(t)$ in both cases are exponential distributed. However they have different means. In figure \ref{strain35} the trajectories of the proportion of infected by the two strains are illustrated. The distribution of the proportion of infected  by strain 1 given that both strains has mayor outbreaks on is illustrated in figure \ref{com35}.

\vfill
\pagebreak

\vfill
\pagebreak

\appendix

\section*{Appendix}\label{Appendix}
We will calculate the right-hand side of equation (\ref{stor}), i.e.
\begin{equation}
A(s)=\int\limits_0^{\infty}\int\limits_0^{\infty}\exp\left(-\lambda p(x-\int\limits_y^{y+x}K(se^{-\alpha u})du)\right)f_{YX}(y,x)dxdy,
\end{equation}
when
\begin{equation}
K(s)=\frac{1}{1+s}.
\end{equation}
Assume that $X$ is exponentially distributed with intensity $\beta$, $Y$ is exponentially distributed with intensity $\delta$, $X$ and $Y$ are independent and 
\begin{equation}
\frac{p\lambda}{\alpha} = 2.
\end{equation}
Elementary calculations yields
\begin{equation}
A(s) = {\rm E}\left(\frac{1+se^{-\alpha(X+Y)}}{1+se^{-\alpha Y}}\right)^2 .
\end{equation}
It is easy to verify that
\begin{equation}
A(0)=1,
\end{equation}
and
\begin{equation}
A'(0)=-\frac{2\delta\alpha}{(\delta +\alpha)(\beta +\alpha)}.
\end{equation}
Further derivations yields
\begin{equation}
A^{(v)}(0) = v!(-1)^v\frac{2\delta\alpha}{(\beta + 2\alpha)(\beta +\alpha)}\frac{\beta +\alpha + v\alpha}{\delta + v\alpha}.
\end{equation}

With  $\delta=\beta +\alpha$ and $\lambda = \beta +2\alpha$, the assumption $p\lambda/\alpha =2$ is satisfied. Furthermore
\begin{equation}
p=\frac{2\delta\alpha}{(\delta +\alpha)(\beta +\alpha)}
\end{equation}
Together with the expressions of the derivatives at $s=0$ a Taylor expansion yields
\begin{equation}
A(s) = 1-p +pK(s).
\end{equation}
Returning to equation (\ref{stor}) this implies that it is solved with $K$ the Laplace transform of an exponential distribution.

\end{document}